\documentstyle[epsf]{mn}

\title[The sdO pulsator J16007+0748]
      {SDSS J160043.6+074802.9: a very rapid sdO pulsator}

\author[P.A. Woudt et. al.]
       {P.A. Woudt$^1$\thanks{pwoudt@circinus.ast.uct.ac.za}, D. Kilkenny$^2$, E. Zietsman$^3$, B. Warner$^1$, 
        N.S. Loaring$^2$, C. Copley$^3$, \newauthor A. Kniazev$^2$, P. V\"ais\"anen$^2$, M. Still$^2$, R.S. Stobie$^2$, E.B. Burgh$^4$, K.H. Nordsieck$^4$,
	\newauthor J.W. Percival$^4$, D. O'Donoghue$^2$ \& D.A.H. Buckley$^2$\\
$^1$Department of Astronomy, University of Cape Town, Private Bag X3, 
Rondebosch 7701, South Africa.\\
$^2$South African Astronomical Observatory, PO Box 9, Observatory 7935,
       South Africa.\\
$^3$National Astrophysics and Space Science Programme, University of 
Cape Town, Rondebosch 7700, South Africa.\\
$^4$Space Astronomy Laboratory, University of Wisconsin, Madison, WI 53706, USA}
\date{Accepted 2006 July 8.  Received 2006 July 7; in original form 2006 May 22}

\begin{document}

\maketitle

\begin{abstract}

We report the serendipitous discovery of the Sloan Digital Sky Survey star,
SDSS J160043.6+074802.9 to be a very rapid pulsator. The variation is dominated
by a frequency near 8380 $\mu$Hz (period = 119.33 s) with a large amplitude
(0.04 mag) and its first harmonic at 16760 $\mu$Hz (59.66 s; 0.005 mag). In
between these frequencies, we find at least another 8 variations with
periods between 62 and 118 seconds and amplitudes between about 0.007 and
0.003 mag; weaker oscillations might also be present. Preliminary
spectrograms from the performance verification phase of the Southern African
Large Telescope indicate that SDSS J160043.6+074802.9 is a spectroscopic binary
consisting of an sdO star and a late-type main-sequence companion. This makes it the first
unambiguous detection of such an sdO star to pulsate, and certainly the first 
found to exhibit multi-frequency variations.
 
\end{abstract}


\section{Introduction}

A decade ago, work on the EC survey (Stobie et al.~1997; 
Kilkenny et al.~1997a) led to the discovery of a new class of pulsating star (Kilkenny et
al.~1997b). These were initially called EC14026 stars after the prototype,
EC14026--2647 -- which is now officially V361 Hya (Kazarovets, Samus \&
Durlevich 2000). EC14026/V361~Hya variables are sdB stars which pulsate with
very short periods (typically $\sim$ 2 to 3 minutes) and which usually have
several (sometimes many) oscillation frequencies. They have surface
temperatures around 28000 $<$ $T_{\rm eff}$ $<$ 40000 K and surface
gravities 5.2 $<$ $\log g$ $<$ 6.1. Independently of the discovery of sdB
pulsations, their existence was predicted on the basis of the presence in
models of low-order and low-degree radial and non-radial $p$-modes driven by
a classical $\kappa$ mechanism associated with an iron opacity peak 
(Charpinet et al.~1996, 1997).

More recently, Green et al.~(2003)  reported the discovery of a separate,
somewhat cooler group of {\it slowly} pulsating sdB stars. These have
periods of around 1 -- 2 hours, which indicates they are probably $g$--mode
pulsators. These slowly pulsating sdB stars are generally cooler than the
fast pulsators, having $T_{\rm eff}$ less than about 27000 K, and are of
somewhat lower surface gravity (log $g$ $\sim$ 5.4) and the two groups are
reasonably (though not totally) separated in a $T_{\rm eff}$/$\log g$ diagram
(see figure 2 of Green et al.~2003). The slow pulsators are usually referred
to as PG1716 stars, after the prototype, PG1716+426.

The way in which sdB stars are formed is not understood, nor is their
relationship to the hotter sdO stars. A number of suggestions have been made
-- delayed helium core flashes or the merger of two helium-rich white dwarf
stars -- and it could be that a number of ways of forming these stars is
possible; they might well be inhomogeneous groups. This certainly seems to
be the case for the sdO stars, which exhibit a wide range of helium
abundances (a recent review has been given by Heber et al. 2006). Unlike the
sdB stars, there are no sdO stars which have been clearly shown to pulsate,
though Rodr\'iguez-L\'opez et al.~(2006) have computed stability analyses of
models for a range of temperatures, gravities and helium abundances which
indicate that sdO stars could pulsate in both $p$- and $g$-modes. These
authors also presented candidate stars as sdO pulsators, but none were
unequivocally demonstrated to vary.

As part of a search by two of us (PW and BW) for new AM CVn stars, a sample
of candidate stars from the Sloan Digital Sky Survey (SDSS; 
Adelman-McCarthy et al.~2006) was selected to
be similar to known stars of this type on the basis of $ugriz$ colours. 
The star discussed in this paper, SDSS J160043.6+074802.9, has colours
derived from SDSS photometry ($u-g = -0.20$ mag, $g-r = -0.20$ mag) which
are identical to that of a newly discovered AM CVn system, 2QZ
J142701.6-012310 (Woudt, Warner \& Rykoff 2005). Of course, selection on the
basis of blue colours does not unequivocally pick out new AM CVn stars and
might include a number of different types of hot object, so a follow-up
programme of ``high-speed'' photometry of likely candidates was carried out.
During the course of this programme, in 2005 May, SDSS J160043.6+074802.9
(which we abbreviate to J16007+0748) was discovered to exhibit pulsations
with a very short period near 120 seconds and indications of a possible
harmonic at half that, making it one of the fastest known pulsators.

The 2000 equatorial co-ordinates are implicit in the full name of the
star; the equivalent galactic co-ordinates are $\ell = 18.65^{\circ};
b = +41.27^{\circ}$. The Sloan Digital Sky Survey gives photometry 
$u = 17.21, g = 17.41, r = 17.61, i = 17.66, z = 17.77$ mag.

The unusually rapid variations of the star prompted us to try for further
observations near the end of the 2005 season; photometric observations to confirm the
variations and to resolve as far as possible all frequencies present, and
spectroscopic data to clarify the nature of the object. In this paper we present
our results to date.

\section{Photometry}

All photometric observations were made with the University of Cape Town CCD
photometer (UCTCCD; O'Donoghue 1995) on the 1.9-m telescope at the Sutherland
site of the South African Astronomical Observatory (SAAO) in July 2005. The
UCTCCD is constructed around a Wright Instruments Peltier-cooled CCD system
utilising a thinned, back-illuminated EEV~P86321/T chip. It is usually used
in frame-transfer mode (one half of the chip is ``masked'' and used as a
storage/slow readout area) so that for exposures longer than about 8
seconds, there is essentially a continuous sequence of observations with no
``dead'' time. (Even faster exposures can be continuous if the CCD is
pre-binned).

On the 1.9-m telescope, the 22$\mu$ pixels of the CCD are equivalent to 0.13
arcseconds at the detector, so that it is normal to use at least 3 x 3
prebinning for optimal data extraction, unless the seeing is better than
about 1 arcsecond. This mode (3 x 3) was used for all the 1.9-m observations
listed in Table 1. All integration times were 10 seconds, which was
considered a reasonable compromise between obtaining as good a
signal-to-noise as possible and having adequate temporal sampling to resolve
the fast variations.

\begin{table}
\centering
\caption{Observation log for J16007+0748 (1.9-m telescope + UCTCCD).}
\vspace{2mm}
\begin{tabular}{|ccccl|}
\hline
Run  &  Date      &   JD &    Run     & Comments  \\
     &  2005      &      &    (hr)    &           \\
\hline
     &  May  &           &            &           \\
S7648& 16/17 &   2453507 &    0.6     & discovery \\
     &  July &           &            &           \\
439  &  5/6  &   2453557 &    1.7     &           \\
443  &  6/7  & ~~~~~3558 &    4.9     &           \\
447  &  7/8  & ~~~~~3559 &    5.3     &  some cloud   \\
451  &  8/9  & ~~~~~3560 &    5.0     &               \\
482  & 15/16 & ~~~~~3567 &    4.9     &  thin cirrus  \\
484  & 16/17 & ~~~~~3568 &    4.4     &  quality poor \\
\hline
\end{tabular}
\end{table}

Reduction of the CCD frames can be performed on-line, which enables the
observer to judge the quality of the observations and to select suitable
stars as local comparisons (to correct for small transparency variations).
Conventional procedures (bias subtraction, flat field correction and so on)
were followed with magnitude extraction being based on the point-spread
function of the DoPHOT program described by Schechter, Mateo \& Saha (1993).

\begin{figure}
\begin{center}
\epsfxsize=70mm
\epsffile{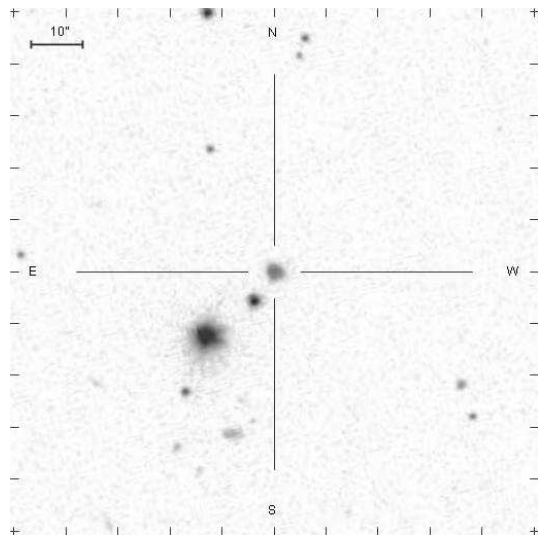}

\caption{Finding chart for J16007+0748; the star is central in the 
chart which is 1.7 x 1.7 arcminutes in size and is taken from the 
Sloan Digital Sky Survey data release 4 (Adelman-McCarthy et al.~2006, 
see http://cas.sdss.org/dr4/en/).}

\label{1600f1}
\end{center}
\end{figure}

\begin{figure}
\begin{center}
\epsfxsize=80mm
\epsffile{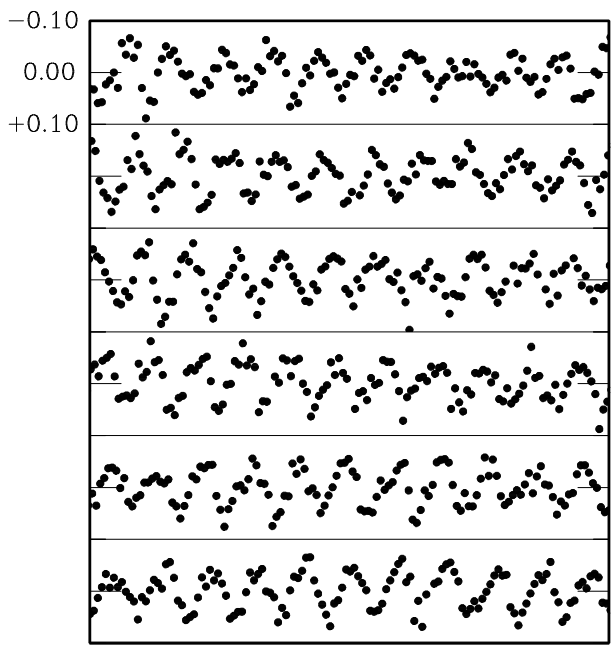}

\caption{A short section of the light curve from the night 2005 July 7/8 
(run 447). The panels read left-to-right and top-to-bottom and each is 
0.015d in length. The  light curve displayed runs from JD 2453559.225 
to .315 and the ordinate scale is magnitudes.}

\label{1600f2}
\end{center}
\end{figure}

The rather small UCTCCD chip (50 x 34 arcsec on the 1.9-m telescope) often
means that it is impossible to get useful comparison stars on the chip.
Fortunately, J16007+0748 has two other stars within about 20 arcsec, one
rather brighter and one significantly fainter (see Fig.~\ref{1600f1}) We have used
the brighter star to correct differentially all observations of the target
star -- to remove rapid transparency variations as well as providing
confidence that the variation is stellar, not atmospheric. Since, in
general, field stars will be quite red, we might expect differential
extinction effects to be significant, and we have further corrected for
atmospheric extinction by removing a second order polynomial from each
night's observations. This means that we might be removing real changes in
stellar brightness on time scales of a few hours but, since we often cannot
distinguish between such effects, this has to be accepted.  A sample of the
differentially corrected light curve is displayed in Fig.~\ref{1600f2}.

\begin{table*}
\centering
\caption{Frequencies and amplitudes extracted one at a time from the 
paired nights and from all nights combined (The column headers give the 
run numbers from Table 1.).}
\vspace{2mm}
\begin{tabular}{|ccccccccc|}
\hline
\multicolumn {2}{|c|}{439 + 443} & \multicolumn {2}{|c|}{447 + 451} & 
\multicolumn {2}{|c|}{482 + 484} & \multicolumn {2}{|c|}{All July 
observations} & \multicolumn {1}{|c|}{} \\
 Freq   & Amp &  Freq   & Amp &  Freq   & Amp &  Freq   & Amp &  P \\
$\mu$Hz & mmag& $\mu$Hz & mmag& $\mu$Hz & mmag& $\mu$Hz & mmag& (s)\\
\hline
~8379.9 & 38.4 & ~8379.8 & 39.3 & ~8379.6 & 38.1 & ~8379.8 & 38.6 & 119.33 \\
~9089.9 & ~7.6 & ~9089.4 & ~6.0 & ~9089.4 & ~7.1 & ~9089.4 & ~6.8 & 110.02 \\
16760.3 & ~4.6 & 16760.0 & ~5.5 & 16749.3 & ~5.0 & 16759.7 & ~5.1 & ~59.66 \\
14188.1 & ~5.2 & 14188.1 & ~5.0 & 14177.6 & ~4.3 & 14188.0 & ~4.7 & ~70.48 \\
        &      & ~8462.7 & ~5.5 & 8472.3  & ~5.4 & ~8475.0 & ~4.4 & 118.00 \\
~9664.5 & ~2.6 & ~9663.8 & ~6.1 & ~9650.4 & ~3.2 & ~9651.9 & ~4.0 & 103.61 \\
13052.5 & ~3.2 &         &      & 13052.4 & ~3.3 & 13052.2 & ~3.5 & ~76.62 \\
15928.3 & ~3.2 & 15927.7 & ~3.8 & 15865.3 & ~3.4 & 15928.4 & ~3.1 & ~62.78 \\  
~9803.1 & ~3.7 & ~9780.1 & ~4.1 &         &      & ~9792.8 & ~2.9 & 102.12 \\
        &      & 13684.6 & ~2.8 & 13660.2 & ~3.9 & 13673.1 & ~2.7 & ~73.14 \\
\hline
\end{tabular}
\end{table*}

\section{Frequency analysis}

The frequency analysis described in this section was carried out using
software which produces Fourier amplitude spectra following the Fourier 
transform method of Deeming (1975) as modified by Kurtz (1985).

\subsection{Individual nights}

Initially, we determined the amplitude spectrum for each night separately.
Even these short data sets, typically 4 to 5 hours, showed up several clear
frequencies (though it is obvious that with periods of 2 minutes or less,
even a run of 4 hours covers much more than 100 cycles). Every one of the
six nights showed the very strong frequency near 8380 $\mu$Hz (119 s) with a
substantial amplitude around 0.039 mag. 
Frequencies around 9090 $\mu$Hz (110
s; 0.007 mag), 14190 $\mu$Hz (70 s; 0.005 mag) and 16760 $\mu$Hz (59 s;
0.005 mag) were also found on every night. The last frequency seems highly
likely to be the first harmonic of the dominant frequency near 8380 $\mu$Hz;
Note that the sidereal drive period of the 74-in telescope is 90 s. 
The dominant frequency and its first harmonic (at 8380 and 16760 $\mu$Hz, respectively)
are therefore intrinsic to the star and not due to the telescope drive error.

In addition, more than half the runs showed apparently significant
frequencies near 13060 $\mu$Hz (73 s), 9660 $\mu$Hz (103 s), 8460 $\mu$Hz
(118 s), 15900 $\mu$Hz (63 s) and 8530 $\mu$Hz (117 s). (We have accepted
amplitudes greater than about 3 times the background noise level -- relaxed
somewhat from the canonical 4 times the background because we do see these
frequencies on so many of the independently analysed nights). These
frequencies are all less than about 0.006 mag in amplitude, though there are
indications that some of them might be variable in amplitude -- they appear
very clearly on some nights and are essentially invisible on others. We
demonstrate this in the next sub-section.

\subsection{Paired nights}

Because of the very rapid nature of the variations and the distribution of
the observations in time (see Table 1), we next determined amplitude spectra 
for data from pairs of consecutive nights. The
results (frequencies and amplitudes) are summarised in the first six columns
of Table 2, where it can be seen that we recover virtually all of the
frequencies mentioned in section 3.1 (we lose the one near 8530 $\mu$Hz) and
find indications that others are split into two components.

\begin{figure}
\begin{center}
\epsfxsize=80mm
\epsffile{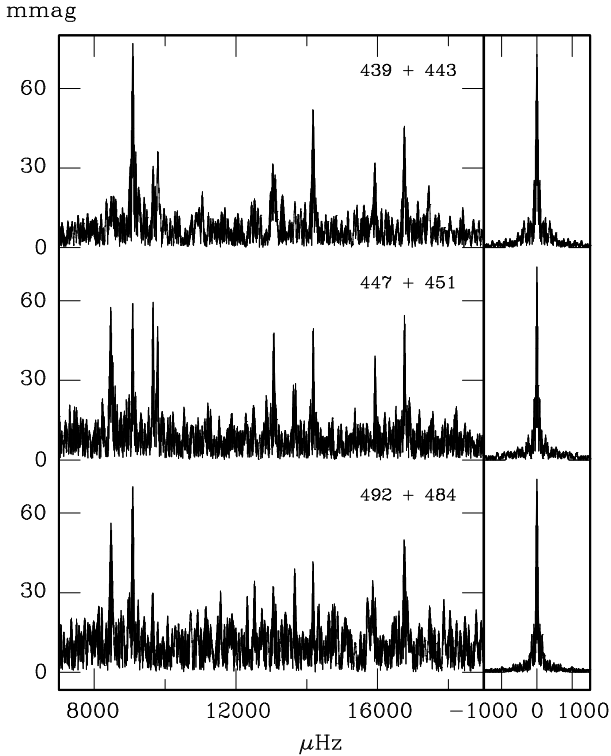}

\caption{Fourier amplitude spectra for paired nights pre-whitened by the 
dominant frequency near 8380 $\mu$Hz (run numbers are shown in the top right
of each plot). The corresponding spectral windows are plotted at the right
with the same frequency scale and with normalised amplitude.}

\label{1600f3}
\end{center}
\end{figure}

It is instructive to examine the Fourier amplitude spectra for the paired
nights, and these are shown in Fig.~\ref{1600f3}. It is clear from this figure that the
frequencies listed in Table 2 are well above the background noise level and
that some of these frequencies are variable in amplitude -- see, for
example, the frequencies between about 8000 and 10000 $\mu$Hz in Fig.~\ref{1600f3},
especially the close pair of frequencies near 9650 and 9800 $\mu$Hz.

\subsection{The full data set}

Because the final set of observations (run 484 -- the night of July 16/17)
appeared rather poor, we determined amplitude spectra for runs 439 to 451
and then runs 439 to 484 inclusively. The results were so similar that in
what follows we have used all of the July data combined.

As in the previously described determinations, we extracted frequencies 
one at a time. These essentially repeated the paired results listed
in Table 2 and so we also include the frequencies determined from the full 
data set in columns 7 and 8 of that Table. 

\begin{figure}
\begin{center}
\epsfxsize=80mm
\epsffile{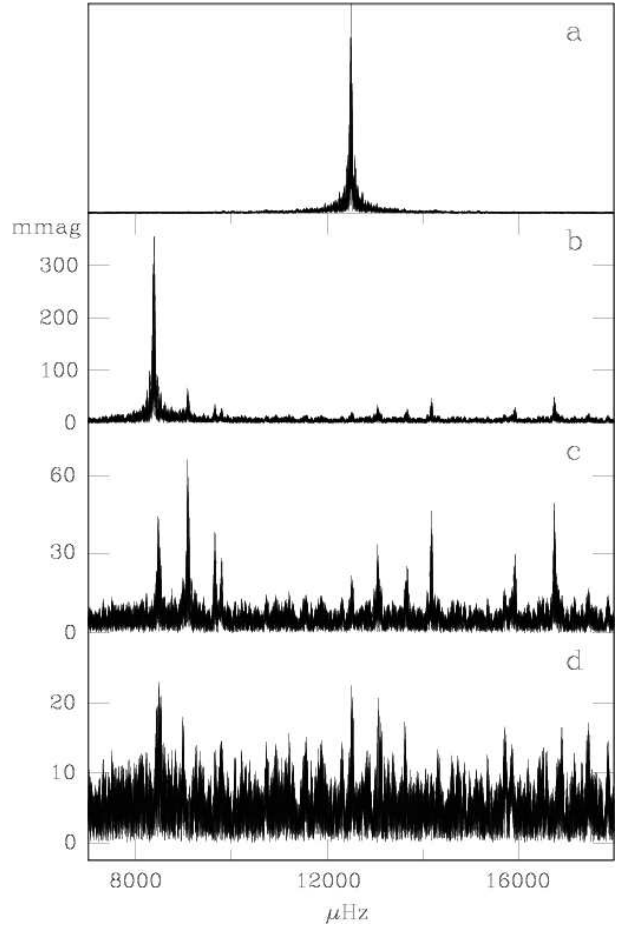}

\caption{Fourier amplitude spectra of the combined 2005 July observations
for J16007+0748. The top panel (a) is the spectral window for the full data
set and panel (b) is the corresponding amplitude spectrum -- the frequency
near 8380 $\mu$Hz dwarfs the other frequencies. Panel (c) shows the
amplitude spectrum pre-whitened by the dominant frequency and panel (d)
shows the spectrum prewhitened by the 10 strongest frequencies; these have
been removed using a simultaneous least squares fit.}

\label{1600f4}
\end{center}
\end{figure}

Given the good inter-agreement between various samplings of the
observations, we have carried out a simultaneous least-squares fit (Deeming
1968) to the ten frequencies which appear to persist in all (or most) of the
solutions, and which are more than four times the general background
noise. The simultaneous solutions are given in Table 3. Fig.~\ref{1600f4} shows the
amplitude spectrum for all the July data; for the spectrum with the dominant
frequency near 8380 $\mu$Hz removed; and with the strongest 10 frequencies
removed.

\begin{table}
\centering
\caption{Frequencies extracted from the full data set for J16007+0748
by simultaneous least-squares fitting. Errors in amplitudes are all around
$\pm$0.4 mmag. Frequencies below the line are less robust.}
\vspace{2mm}
\begin{tabular}{|cccccc|}
\hline
         &  Freq    & $\sigma$ &  Amp   & Period \\
         & $\mu$Hz  &  $\pm$   &  mmag  &   (s)  \\
\hline
$f_1$    &  ~8379.82 & 0.01 & 39.8 & 119.33  \\
$f_2$    &  ~9089.40 & 0.03 & ~6.8 & 110.01  \\
$f_3$    &  16759.68 & 0.03 & ~5.1 & ~59.67  \\
$f_4$    &  14187.95 & 0.04 & ~4.8 & ~70.48  \\
$f_5$    &  ~8462.14 & 0.04 & ~4.8 & 118.17  \\
$f_6$    &  ~9651.90 & 0.04 & ~4.0 & 103.61  \\
$f_7$    &  13052.22 & 0.05 & ~3.4 & ~76.61  \\
$f_8$    &  15940.01 & 0.06 & ~3.0 & ~62.74  \\
$f_9$    &  ~9804.37 & 0.06 & ~2.8 & 102.00  \\
$f_{10}$ &  13673.08 & 0.07 & ~2.6 & ~73.14  \\
\hline
$f_{11}$ &  ~8477.90 & 0.07 & ~2.5 & 117.95  \\
$f_{12}$ &  12526.81 & 0.07 & ~2.4 & ~79.83  \\
$f_{13}$ &  13071.88 & 0.07 & ~2.2 & ~76.50  \\
$f_{14}$ &  ~8516.42 & 0.08 & ~2.2 & 117.42  \\   
\hline
\end{tabular}
\end{table}

In Fig.~\ref{1600f4}d it does appear that there are still peaks well above the
background noise. The rms scatter in the background over the whole range of
frequency shown in the figure is about 0.5 to 0.6 mmag and our
criterion for ``reality'' of frequency peaks is that they should be
more than four times the noise level 
We have therefore extracted a further four frequencies from the residual
amplitude spectrum in Fig.~\ref{1600f4}d which are close to 2.2 mmag in amplitude and
these are listed below the line in Table 3. They are less secure than $f_1$
to $f_{10}$ and it should be mentioned that two of them, $f_{11}$ and
$f_{14}$ are close to the very strong frequency, $f_1$, and might be an
artefact of imperfect extraction of that frequency. With 14 frequencies
removed, even the most enthusiastic frequency-extractor would be hard put to
find anything in our observations significantly different from noise.

\begin{figure}
\begin{center}
\epsfxsize=80mm
\epsffile{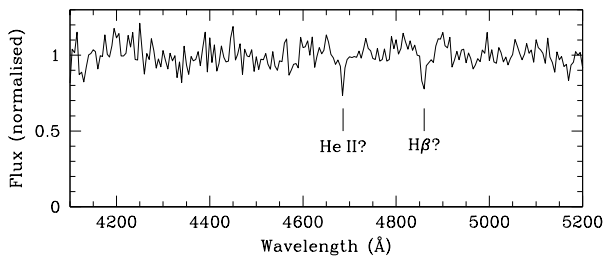}

\caption{Spectrogram of J16007+0748 from six co-added exposures 
totalling 8400 seconds and made with the SAAO 1.9-m telescope and Cassegrain
spectrograph. The displayed spectrogram was smoothed by applying a running
mean to the original; this reduced the pixel scale from 1.1 to 5 \AA/pixel.}

\label{1600f5}
\end{center}
\end{figure}

\section{Spectroscopy}

The fact that J16007+0748 was selected as a very hot object and the nature
of the observed photometric variations indicated that the object was likely
to be a rapidly pulsating sdB star. We were therefore keen to get
spectroscopic observations to determine the nature of the star.

\subsection{1.9-m spectroscopy}

Although J16007+0748 is fainter than 17 mag (see Section 1) and therefore right
at the faintness limit of the spectrograph on the SAAO 1.9-m telescope, we used the
earliest time available to us on the SAAO 1.9-m telescope with the Cassegrain
spectrograph. Unfortunately, this was near the end of the season for
J16007+0748, but we obtained two 15 -- 20 minute exposures at the start
of each night between 1/2 and 3/4 September 2005, giving us a total of six
(very noisy) spectrograms and a total integration time of 8400 seconds.

All spectrograms were obtained with grating 6, which has a spectral range of
about 3500 -- 5300 \AA \, at a resolution of around 4 \AA. A rather wide
slit was used (300 $\mu$, equivalent to 1.8 arcsec on the sky) to maximise
light throughput. The final co-added spectrogram is shown in Fig.~\ref{1600f5}; this has
been smoothed and the blue end truncated because the noise was so high.
There is little that can be confidently deduced from Fig.\ref{1600f5}; it is probable
that He\,II 4686 and H(/He\,II?) at 4861 \AA \, are present, but the evidence is
not overwhelming. However, if He\,II is present at comparable strength to the
$\lambda$4861 line, then the star can not be of sdB type.

\begin{figure*}
\begin{center}
\epsfxsize=160mm

\epsffile{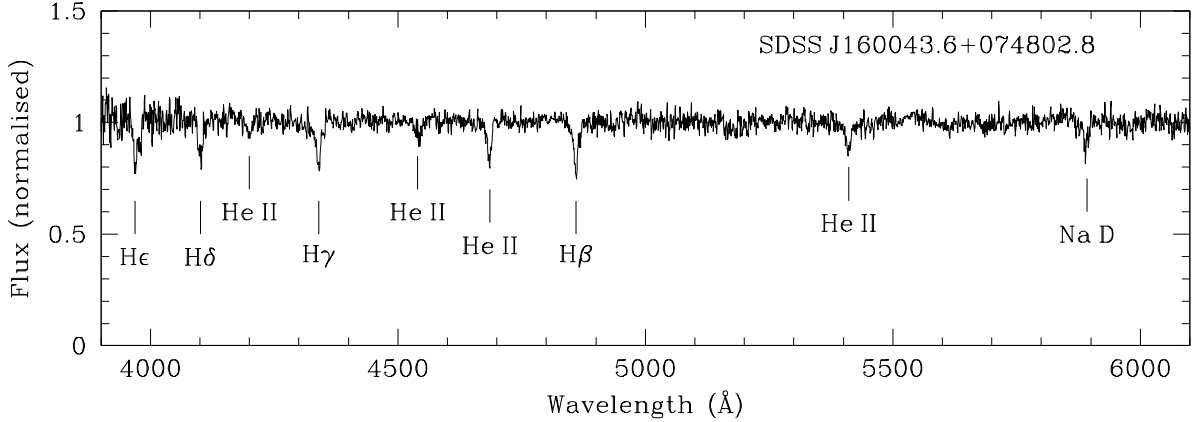}

\caption{Combined SALT spectrogram of J16007+0748 obtained with a total exposure
time of 1800 seconds. The displayed spectrogram is an average of three 600-s exposures
taken on April 4 and June 5 2006, respectively. The absorption lines are marked and labelled.}

\label{1600f6}
\end{center}
\end{figure*}

\begin{table*}
\centering
\caption{Differences between observed ($\lambda_{obs}$) and rest 
($\lambda_{lab}$) wavelengths, and heliocentric velocities derived for the strong lines in Fig.~\ref{1600f6}.}
\vspace{2mm}
\begin{tabular}{|lcrrrr|}
\hline
 &   &  \multicolumn{2}{c}{4 April 2006} & \multicolumn{2}{c}{5 June 2006} \\  
 &  $\lambda_{lab}$ (\AA) & $\lambda_{obs}$ (\AA) & $v_{\rm hel}$ (km s$^{-1}$) & $\lambda_{obs}$ (\AA) & $v_{\rm hel}$ (km s$^{-1}$) \\
\hline
H$\epsilon$ &  3970.07   &    3969.67 $\pm$ 0.29  &   --13 $\pm$ 22   &   \multicolumn{2}{c}{not in spectral range} \\
H$\delta$   &  4101.73   &    4101.10 $\pm$ 0.25  &   --29 $\pm$ 18   &   \multicolumn{2}{c}{not in spectral range} \\
He\,II      &  4199.83   &    \multicolumn{2}{c}{low signal-to-noise, no clear line} & 4199.38 $\pm$ 0.18  &   --40 $\pm$ 13 \\
H$\gamma$   &  4340.46   &    4339.47 $\pm$ 0.25  &   --51 $\pm$ 17   &    4339.13 $\pm$ 0.13  &   --100 $\pm$ \, 8\\
He\,II      &  4541.59   &    \multicolumn{2}{c}{in gap between two CCDs in mosaic}  &    4539.51 $\pm$ 0.16  &  --146 $\pm$ 11  \\
He\,II      &  4685.71   &    4685.36 $\pm$ 0.24  &    --5 $\pm$ 15   &    4684.42 $\pm$ 0.14  &   --91 $\pm$ \, 9 \\
H$\beta$    &  4861.33   &    4860.57 $\pm$ 0.24  &   --29 $\pm$ 15   &    4859.53 $\pm$ 0.13  &  --119 $\pm$ \, 8 \\
He\,II      &  5411.53   &    5410.87 $\pm$ 0.31  &   --19 $\pm$ 17   &    5409.37 $\pm$ 0.16  &  --128 $\pm$ \, 9 \\
Na\,D       &  5889.95   &    (5889.33 $\pm$ 0.26)  &   (--14 $\pm$ 13)   &   \multicolumn{2}{c}{blended}         \\
\hline
\end{tabular}
\end{table*}

\subsection{SALT spectroscopy}

The Southern African Large Telescope (SALT) is currently in the
``performance verification'' phase where engineering needs are predominant,
but we were fortunate to be able to get three 600-second spectrograms of good
signal-to-noise in April and June 2006. Three additional 600-seconds spectrograms 
were obtained in the same period in rather too poor conditions (bright Moon plus cloud) 
to be of much further use.
 
The spectrograms were taken with the Robert Stobie Spectrograph (previously
called the Prime Focus Imaging Spectrograph) in single-slit spectroscopy 
mode using the PG1300 grating at a grating angle of 18.5$^{\circ}$ in April 2006
(corresponding to a camera articulation angle of 37$^{\circ}$ in
a standard Littrow configuration) and at a grating angle of 19.62$^{\circ}$ in June 2006. 
A 1.5-arcsec slit was used, providing a central wavelength of 4880 {\AA} (April 2006) and
5170 {\AA} (June 2006), respectively, and a resolution of 3.9 \AA.

The flat-fielded, wavelength-calibrated SALT spectrogram is shown in Fig.~\ref{1600f6}.
This spectrum is an average of the three spectra obtained in April and June 2006.
The He\,II $\lambda$4686 line and the Pickering series of He\,II at $\lambda$5411, $\lambda$4542 and
$\lambda$4200 are clearly present in absorption. 
The Balmer series lines H$\beta$ to H$\epsilon$ are present,
but these are probably blended with Pickering lines of He\,II. 

Table 4 shows a
comparison of the measured wavelengths ($\lambda_{obs}$) with the rest
wavelengths ($\lambda_{lab}$). The He\,II $\lambda$4686 and $\lambda$5411
lines measured in the spectrogram obtained in April 2006, 
which should be relatively unblended, show displacements equivalent
to --12 km s$^{-1}$ (corrected to heliocentric velocities) whereas the hydrogen
Balmer series lines show substantially higher blue shifts -- consistent with
the Balmer lines being blended with He\,II Pickering lines. Note, however, that 
this difference in heliocentric velocity is of the same order of magnitude as the 
errors in the heliocentric velocities. These errors are derived
from the standard deviation in the wavelength-calibration solution using
a Cu-Ar calibration spectrogram ($\sigma \sim 0.24$ {\AA} in April 2006 and 
0.12 {\AA} in June 2006, respectively)
and from repeated measurements of individual lines; typically each line was 
measured 5 times. 

No He\,I lines (such
as the typically strong features near 4471 and 4026 \AA) are detected,
ruling out types sdB and sdOB. (He\,II is sometimes detected in these
relatively cooler stars, but it is much weaker than the Balmer lines or He\,I). 
The spectrum is clearly that of a classical sdO star (see, for example,
Moehler et al. 1990).

Based on the April 2006 SALT spectrogram of J16007+0748, the referee suggested the
possibility of the star being a spectroscopic binary consisting of an sdO and sdB star (where the
pulsations could be due to the sdB star instead). To explore this possibility further,
the spectrograms in June 2006 were obtained. Apart from the considerable shift in heliocentric
velocities of the measured lines over a 2-month period (see Table 4), 
the Na\,D lines $\lambda$5889, 5896 (marginally resolved) appeared clearly in the spectrum. The large
equivalent width of the Na\,D lines and the high Galactic latitude of J16007+0748 ($b = +41^{\circ}$) 
rules out the possibility of interstellar Na\,D absorption lines.
This suggests that J16007+0748 is indeed a spectroscopic binary, but one consisting of an
sdO star and a late-type main sequence star and appears to rule out an sdB component.

We therefore suggest that J16007+0748 is the first clearly established sdO
pulsator. We note that Rodr\'iguez-L\'opez et al. (2006) have presented
candidate stars as sdO pulsators, but these are of very low amplitude and
remain to be confirmed. At the least, we have demonstrated unequivocal
multi-frequency variations in an sdO star for the first time.

\section{Discussion}

We have shown SDSS J160043.6+074802.9 to be a spectroscopic binary consisting of
an sdO and late-type main sequence companion.
In addition we have shown it to have at least 10 modes of variation
in the range 60 to 120 s (16760 to 8389 $\mu$Hz). Given this array
of frequencies, it seems highly likely that the source of the variation is
pulsation in the sdO star.

Until we obtained the SALT spectrogram in 2006 April, we had assumed the
object would turn out to be a rapidly pulsating sdB star (= EC14026 star =
V361 Hya star; all unlovely names). These generally show:

\begin{itemize} 

\item Very rapid variations, with typical periods $\sim$ 2 -- 8 minutes.

\item Complex variations, in that most show several frequencies -- and
some show a large number ($>$ 40).

\item  Additional complexity, in that many stars have frequencies with
changing amplitudes -- indicating amplitude variations or that we 
might be observing the beating of as yet unresolved frequencies.

\end{itemize}

J16007+0748 exhibits all these properties. This is presumably because the
sdO stars -- like the sdB stars -- are evolved, low-mass ($\sim$ 0.5
M$_{\odot}$) stars although, of course, the sdO stars are hotter.
Rodr\'iguez-L\'opez et al. (2006) have presented a synopsis of stability
analyses of model sdO stars. They find three frequency regions where
pulsations are stable; two regions where modes have a tendency to
instability -- 500 to 2000 $\mu$Hz corresponding to high radial order
$g$-modes and 9000 to 12000 $\mu$Hz corresponding to low radial order
$p$-modes -- and a region of high stability from about 2000 -- 9000 $\mu$Hz.
The observed variations in J16007+0748 lie between 8000 and 16000 $\mu$Hz
and so are of the right order and would suggest $p$-modes -- as is the case
with the rapidly pulsating sdB stars. Of course, the models were calculated
for specific temperatures and gravities and we do not yet have
determinations of these for this star, but the broad agreement is promising.
Substantially higher S/N spectrograms (at higher spectral resolution to
resolve the H/He\,II blended lines) are needed to do a full spectral
analysis on this star and compare its properties ($T_{\rm eff}$, $\log g$
and chemical composition) with other sdOs and with the already known classes
of hot pulsators (the PG\,1159 stars, Werner 2001).

The substantial shift in heliocentric velocities observed over a 2-month period
of order $\sim$ 80 km s$^{-1}$ (see Table 4), suggests that J16007+0748 is a short-period binary and further 
spectroscopic observations are needed to resolve the spectroscopic period of this binary.
We tentatively constrain the spectral type of the companion to late F or early G based
on the presence of the Na\,D line, and the absence of Ca\,I $\lambda$6162 and the blend of lines
near 6497 {\AA} (see the spectral atlas of Allen \& Strom 1995). The latter is based on a
20-min spectrogram obtained with SALT in marginal conditions (bright moon)
in the red part using the G0900 grating covering $\sim$ 5000--8000 {\AA}.

Further evidence for the binary nature of J16007+0748 comes from the observed SDSS
colours (see Section 1). The $u-g$ and $r-i$ colours of J16007+0748 deviate significantly
from the well-defined concentration of subdwarfs in the colour-colour diagram of white dwarfs and
subdwarfs identified in the SDSS (see figure 2 of Kleinman et al.~2004).

\section*{Acknowledgements}

PAW kindly acknowledges financial support from the University of 
Cape Town and the National Research Foundation.
We thank the referee, Stefan Dreizler, for very useful comments regarding
the possible binary nature of J16007+0748.
Funding for the creation and distribution of the SDSS Archive has
been provided by the Alfred P. Sloan Foundation, the Participating
Institutions, the National Aeronautics and Space Administration, the
National Science Foundation, the U.S. Department of Energy, the Japanese
Monbukagakusho, and the Max Planck Society. The SDSS Web site is
http://www.sdss.org/

The SDSS is managed by the Astrophysical Research Consortium (ARC) for the
Participating Institutions. The Participating Institutions are The
University of Chicago, Fermilab, the Institute for Advanced Study, the Japan
Participation Group, The Johns Hopkins University, the Korean Scientist
Group, Los Alamos National Laboratory, the Max-Planck-Institute for
Astronomy (MPIA), the Max-Planck-Institute for Astrophysics (MPA), New
Mexico State University, University of Pittsburgh, Princeton University, the
United States Naval Observatory, and the University of Washington.


\end{document}